\documentclass[aps,prl,twocolumn,letterpaper,groupedaddress]{revtex4-1}

\usepackage{color}

\usepackage{float}
\usepackage{wrapfig}
\usepackage{caption}
\usepackage{graphicx}
\usepackage{pgfplots}
\usepackage{subcaption}
\usepackage{tikz,tikz-3dplot}
\tdplotsetmaincoords{80}{45}
\tdplotsetrotatedcoords{-90}{180}{-90}

\tikzset{surface1/.style={draw=blue!70!black, fill=blue!40!white, fill opacity=.6}}
\tikzset{surface2/.style={draw=red!70!black, fill=red!40!white, fill opacity=.6}}
\tikzset{surface3/.style={draw=green!70!black, fill=green!40!white, fill opacity=.6}}

\usetikzlibrary{intersections,fadings,decorations.pathreplacing}

  \pgfplotsset{
        compat=1.8}

\def \be {\begin{equation}}
\def \ee {\end{equation}}
\def \bea {\begin{eqnarray}}
\def \eea {\end{eqnarray}}
\newcommand{\eqref}[1]{(\ref{#1})}

\begin{document}

\title{Well-posed formulation of scalar-tensor effective field theory}



\author{Aron D. Kovacs and Harvey S. Reall}
\email[]{adk42@cam.ac.uk, hsr1000@cam.ac.uk}
\affiliation{Department of Applied Mathematics and Theoretical Physics, University of Cambridge, Wilberforce Road, Cambridge CB3 0WA, United Kingdom}


\date{\today}

\begin{abstract}
Effective field theory provides a way of parameterizing strong-field deviations from General Relativity that might be observable in the gravitational waves emitted in a black hole merger. To perform numerical simulations of mergers in such theories it is necessary that the equations be written in a form that admits a well-posed initial value formulation. We study gravity coupled to a scalar field including the leading (4-derivative) effective field theory corrections. We introduce a new class of ``modified harmonic" gauges and gauge-fixed equations of motion, such that, at weak coupling, the equations are strongly hyperbolic and therefore admit a well-posed initial value formulation. 
\end{abstract}


\maketitle

\section{Introduction}

The detection of gravitational waves from black hole (BH) mergers \cite{Abbott:2016blz} is an opportunity to perform the first precision tests of General Relativity (GR) in a strong field, highly dynamical regime. To do this, we need theoretical templates for how a deviation from GR would affect the gravitational waves produced in a BH merger. Producing such templates requires numerical relativity simulations of BH mergers in theories that modify GR in some way. But there are two problems with this (see e.g. \cite{Yunes:2016jcc}). First: which theory should be simulated? Many theories of modified gravity have been proposed. Second: to perform numerical simulations, it is essential that the the theory is written in a form that admits a {\it well-posed initial value problem}. This means that, given suitable initial data, there exists a unique solution of the equations of motion that depends continuously on the data.

Effective field theory (EFT) provides a possible solution to the first problem \cite{Endlich:2017tqa}. Without a preferred candidate for whatever ``UV physics" modifies GR, we can parameterize our ignorance using the EFT methodology of adding to the GR Lagrangian all possible higher derivative terms and then using observations to constrain the coefficients of these terms. This provides a nice way of parameterizing small strong-field deviations from GR. The accuracy to which one has tested GR can be quantified by how small one has constrained the coefficients of the leading higher derivative terms to be. Unfortunately, if one tries to do this for vacuum gravity, one runs into the second problem. This is because, after field redefinitions, the leading higher derivative corrections to vacuum GR start at $6$ derivatives \cite{Endlich:2017tqa}. The equation of motion now involves higher than second derivatives of the metric and therefore is unlikely to admit a well-posed initial value problem. (See \cite{Allwright:2018rut} for discussion of this problem.)

If one includes matter then one can do better. The simplest case is GR minimally coupled to a scalar field. Following the EFT philosophy, one adds all possible higher derivative terms to the action. Assuming a parity symmetry, field redefinitions can be used to bring the action to the form \cite{Weinberg:2008hq}
\be
\label{4dST}
 S = \int \frac{ d^4 x \sqrt{-g}}{16\pi G} \left( - V(\phi)+ R + X +  \alpha(\phi) X^2 +  \beta(\phi) {\cal L}_{\rm GB}  \right)
\ee
where we have neglected terms with $6$ or more derivatives, $V,\alpha,\beta$ are arbitrary functions, $X =-(1/2) g^{\mu\nu} \partial_\mu \phi \partial_\nu \phi$ and ${\cal L}_{\rm GB}$ is the Euler density associated to the Gauss-Bonnet invariant
\be
 {\cal L}_{GB}=\frac14 \delta^{\mu_1\mu_2\mu_3\mu_4}_{\nu_1\nu_2\nu_3\nu_4}R_{\mu_1\mu_2}{}^{\nu_1\nu_2}R_{\mu_3\mu_4}{}^{\nu_3\nu_4}.
\ee
The coupling of the scalar field to ${\cal L}_{GB}$ implies that spacetime curvature is a source for the scalar field, which must therefore be non-zero near a BH. This may cause observable deviations from GR in a BH merger. If one imposes an additional symmetry that the equations of motion are invariant under shifts in $\phi$ then $V$ and $\alpha$ are constants and $\beta=\lambda\phi$ where $\lambda$ is a constant. The dimensionful constants $\alpha,\lambda$ then set a scale for UV physics.

EFT reasoning implies that the theory \eqref{4dST} is also relevant to cosmology e.g. in early Universe inflation \cite{Weinberg:2008hq}. 

Remarkably, the equations of motion of \eqref{4dST} are second order in derivatives. Hence it is possible that this theory admits a well-posed initial value problem. Note that neglect of terms in the action with $6$ or more derivatives is justified only in a regime in which spacetime curvature and scalar field derivatives are small compared to the UV length scales introduced by coupling constants associated with the higher derivative corrections. Generically, this implies that the $4$-derivative corrections to the equations of motion must also be small compared to the leading $2$-derivative terms. We refer to this as the {\it weakly coupled} regime. It is only in this regime that we can trust EFT. Weak coupling is compatible with strong-field BH dynamics, as long as the size of the BHs is large compared to the UV length scales. 

Establishing well-posedness requires finding a ``good gauge" for the equations of motion and a good way of performing the gauge fixing. The simplest choice for GR is harmonic gauge, but it has been shown that this doesn't work for \eqref{4dST}: the initial value problem is not well-posed even at weak coupling \cite{Papallo:2017qvl,Papallo:2017ddx}. This means that numerical simulations of theories of the above type have been restricted either to spherical symmetry \cite{Ripley:2019hxt,Ripley:2019irj,Ripley:2019aqj} or to solving the equations perturbatively (in $\lambda$ for the case $\alpha=0$, $\beta(\phi) = \lambda \phi$) \cite{Witek:2018dmd,Okounkova:2019zep,Okounkova:2020rqw}. The latter approach can suffer from small effects gradually accumulating over time, leading to a breakdown of perturbation theory in situations when the EFT should be valid. A well-posed formulation of the equations should be able to handle such secular effects \cite{Flanagan:1996gw}.

In this Letter, we will introduce modifications of the harmonic gauge condition and gauge-fixing procedure used in GR. We will use these to define gauge-fixed equations of motion for \eqref{4dST} and explain why these equations admit a well-posed initial value problem at weak coupling. Our formulation opens up the possibility of performing numerical simulations of black hole mergers in this theory without resorting to perturbation theory.

We follow the conventions of \cite{Wald:1984rg}. Indices $\mu,\nu,\ldots$ run from $0$ to $3$, Indices $i,j,\ldots$ run from $1$ to $3$.

\section{Modified harmonic gauge}

In a spacetime $(M,g)$, introduce two auxiliary (inverse) Lorentzian metrics $\tilde{g}^{\mu\nu}$ and $\hat{g}^{\mu\nu}$ such that the causal cone of $g^{\mu\nu}$ (i.e. the set of timelike or null covectors) is strictly inside the causal cone of $\tilde{g}^{\mu\nu}$, and the latter is strictly inside the causal cone of $\hat{g}^{\mu\nu}$ (Fig. 1(a)). Raising and lowering of indices will always be performed using the physical metric. We write the inverses of $\tilde{g}^{\mu\nu}$ and $\hat{g}^{\mu\nu}$ as $(\tilde{g}^{-1})_{\mu\nu}$ and $(\hat{g}^{-1})_{\mu\nu}$. The causal cone of $(\hat{g}^{-1})_{\mu\nu}$ lies strictly inside that of $(\tilde{g}^{-1})_{\mu\nu}$, which lies strictly inside that of $g_{\mu\nu}$ (Fig. 1(b)). These relations imply that a surface that is spacelike w.r.t. $g_{\mu\nu}$ is also spacelike w.r.t. the other two metrics. They also imply that $D(\Sigma) \subset \hat{D}(\Sigma)$ where $D(\Sigma)$ and $\hat{D}(\Sigma)$ are the domains of dependence of a partial Cauchy surface $\Sigma$ defined in the usual way \cite{Wald:1984rg} w.r.t. the metrics $g_{\mu\nu}$ and $(\hat{g}^{-1})_{\mu\nu}$ respectively.

\begin{figure}[H]
\centering
\begin{subfigure}{0.5\linewidth}
\centering
 \includegraphics[scale=1.0]{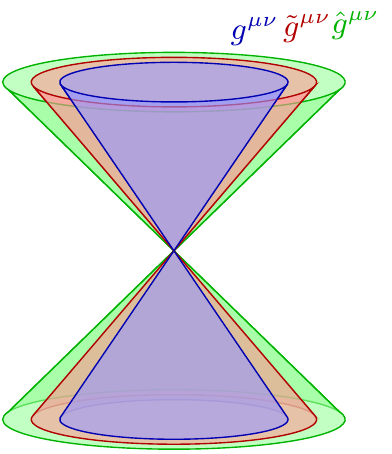}
\caption{}
\label{fig:cones1}
\end{subfigure}%
\begin{subfigure}{0.5\linewidth}
\centering
 \includegraphics[scale=1.0]{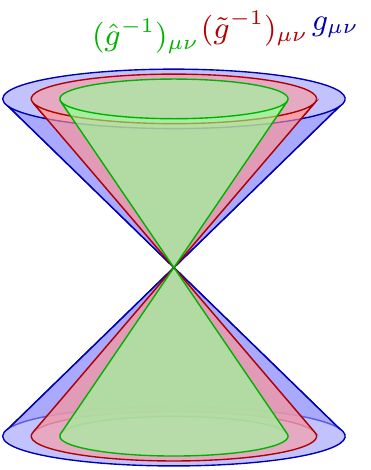}
\caption{}
\label{fig:cones2}
\end{subfigure}
\caption{(a) Cotangent space, showing the
null cones of $g^{\mu\nu}$, $\tilde{g}^{\mu\nu}$ and $\hat{g}^{\mu\nu}$. 
(b) Tangent space, showing the null cones of $g_{\mu\nu}$, $(\tilde{g}^{-1})_{\mu\nu}$ and $(\hat{g}^{-1})_{\mu\nu}$.}
\label{fig:cones}
\end{figure}

Our modified harmonic gauge condition on the coordinates $x^\mu$ is $H^\mu=0$ where
\be
 H^\mu \equiv \tilde{g}^{\nu\rho}\nabla_\nu \nabla_\rho x^\mu = -\tilde{g}^{\nu \rho} \Gamma^\mu_{\nu\rho}
\ee
Given initial data for the coordinates $x^\mu$ on a surface $\Sigma$ spacelike w.r.t. $g^{\mu\nu}$ (and hence also w.r.t. $\tilde{g}^{\mu\nu}$), this equation can be solved to construct coordinates in the same way as in harmonic gauge GR \cite{Wald:1984rg}.

We now let
\be
 E^{\mu\nu} = -\frac{16 \pi G}{\sqrt{-g}} \frac{\delta S}{\delta g_{\mu\nu}} \qquad \qquad E_\phi = -\frac{16 \pi G}{\sqrt{-g}} \frac{\delta S}{\delta \phi}
\ee
The diffeomorphism invariance of our theory implies that these satisfy the Bianchi identity
\be
\label{bianchi}
\nabla^\mu E_{\mu\nu}-E_\phi\nabla_\nu\phi=0.
\ee
The equations of motion of \eqref{4dST}, before gauge fixing, are
\be
\label{eofm}
E^{\mu\nu} = E_\phi=0
\ee

We now define
\be\label{gauge_fixing}
 E_{\rm mhg}^{\mu\nu} = E^{\mu\nu} + \hat{P}_\alpha{}^{\beta \mu \nu} \partial_\beta H^\alpha
\ee
where $
 \hat{P}_\alpha{}^{\beta \mu \nu} = \delta_\alpha^{(\mu} \hat{g}^{\nu)\beta}- \frac{1}{2} \delta_\alpha^\beta \hat{g}^{\mu\nu}$.
Our modified harmonic gauge equations of motion are then
\be
\label{def_har}
 {E}_{\rm mhg}^{\mu\nu}=0 \qquad \qquad E_\phi=0
\ee
If we set $\tilde{g}^{\mu\nu}=\hat{g}^{\mu\nu}=g^{\mu\nu}$ then these reduce to the usual harmonic gauge equations of motion. The latter do not admit a well-posed initial value problem \cite{Papallo:2017qvl,Papallo:2017ddx}. The reason for this can be traced to the fact that, in harmonic gauge, unphysical ``gauge-condition violating" modes travel at the same speed as ``pure gauge" modes. Choosing $\tilde{g}^{\mu\nu}$ and $\hat{g}^{\mu\nu}$ as explained above eliminates this degeneracy. It also ensures that the ``fastest" modes are the physical modes. 

We will now sketch the proof that the initial value problem for \eqref{def_har} is well-posed. A full proof will appear in \cite{newpaper}.

A slight modification of the usual argument for harmonic gauge GR \cite{Wald:1984rg} can be used to prove that \eqref{def_har} propagates the gauge condition. Given a solution $(M,g,\phi)$ of \eqref{def_har}, equation \eqref{bianchi} implies
\be
\label{Heq}
 0 = \nabla_\nu E_{\rm mhg}^{\mu\nu} = \frac{1}{2} \hat{g}^{\alpha \beta} \partial_\alpha \partial_\beta H^\mu + \ldots
\ee
where the ellipsis denotes terms linear in first derivatives of $H^\rho$. Let $\Sigma \subset M$ be a surface that is spacelike w.r.t. $g^{\mu\nu}$ and hence spacelike w.r.t. $\hat{g}^{\mu\nu}$. Equation \eqref{Heq} admits a well-posed initial value problem for initial data $H^\mu$ and $\hat{g}^{\nu\rho}n_\nu \partial_\rho H^\mu$ prescribed on $\Sigma$ (where $n$ is the unit normal of $\Sigma$ w.r.t. $g$). Hence any solution of \eqref{def_har} for which $H^\mu$ and its normal derivative vanish on $\Sigma$ will have $H^\mu\equiv 0$ in $\hat{D}(\Sigma)$ and therefore satisfy \eqref{eofm} in $\hat{D}(\Sigma)$.

Initial data is a quintuple $(\Sigma,h_{ij},K_{ij},\Phi,\Psi)$ where $\Sigma$ is a $3$-manifold and, in a coordinate chart $x^i$ on $\Sigma$, $h_{ij}$ is a Riemannian metric on $\Sigma$, $K_{ij}$ a symmetric tensor and $\Phi,\Psi$ are functions on $\Sigma$ specifying the scalar field and its normal derivative on $\Sigma$. These must satisfy the constraint equations arising from \eqref{eofm}. We perform a $3+1$ split of $g_{\mu\nu}$, with coordinates $x^\mu=(x^0,x^i)$ and using the usual lapse function and shift vector \cite{num_rel}. This ensures that surfaces of constant $x^0$ are spacelike w.r.t. $g^{\mu\nu}$ and hence also w.r.t. the other two metrics. At $x^0=0$ the choice of lapse and shift is arbitrary. Given such a choice, the initial values of $g_{ij}$ and $\partial_0 g_{ij}$ are chosen so that the surface $x^0=0$ has induced metric $h_{ij}$ and extrinsic curvature $K_{ij}$. The initial value of $\phi$ is $\Phi$ and the initial value of $\partial_0 \phi$ is chosen so that $n \cdot \partial \phi = \Psi$. The time derivatives of the lapse and shift at $x^0=0$ are uniquely specified by demanding $H^\mu =0$ at $x^0=0$. This implies $\partial_i H^\mu=0$ at $x^0=0$. Evaluating the $0\mu$ components of \eqref{def_har} and using the constraint equations $E^{0\mu}=0$ gives $\partial_0 H^\mu=0$ at $x^0=0$. Given a solution $(M,g,\phi)$ of \eqref{def_har} arising from this initial data, we identify $\Sigma$ with the surface $x^0=0$, and the argument above shows that $H^\mu \equiv 0$ in $\hat{D}(\Sigma)$ hence \eqref{eofm} is satisfied in $D(\Sigma) \subset \hat{D}(\Sigma)$. 

Sufficient conditions for well-posedness of the initial value problem for \eqref{def_har} are that the equations are {\it strongly hyperbolic} and the initial data is prescribed on a surface that is {\it non-characteristic} \cite{Taylor91}. See \cite{Papallo:2017qvl} or \cite{Sarbach2012} for definitions of these terms. 

The principal symbol of \eqref{def_har} is calculated by linearizing around an arbitrary ``background" field configuration and making the replacements $\partial_\mu \partial_\nu \delta g_{\rho\sigma} \rightarrow \xi_\mu \xi_\nu t_{\rho\sigma}$ and $\partial_\mu \partial_\nu \delta \phi\rightarrow \xi_\mu \xi_\nu \psi$ where $\xi_\mu$ is an arbitrary covector and $t_{\mu\nu}$ is symmetric. We combine $t_{\mu\nu}$ and $\psi$ into a vector $T_I=(t_{\mu\nu},\psi)^T$ where indices $I,J, \ldots$ refer to a basis for the 11-dimensional space of such vectors. The principal symbol of \eqref{def_har} is an $11\times 11$ matrix ${\cal P}^{IJ}(\xi)={\cal P}^{IJ \mu\nu} \xi_\mu \xi_\nu$ where ${\cal P}^{IJ \mu\nu}$ depends on the background metric, Riemann tensor, and up to two derivatives of the background $\phi$ field. The covector $\xi_\mu$ is {\it characteristic} if there exists $T_I \ne 0$ such that
\be
\label{characteristic}
 {\cal P}^{IJ}(\xi) T_J = 0
\ee
equivalently $\det {\cal P}(\xi)=0$. A characteristic covector corresponds to the wavevector of a high frequency wave solution of \eqref{def_har}, with polarization $T_I$.

As discussed above, by writing our initial metric in $3+1$ (lapse-shift) form with coordinates $(x^0,x^i)$ we ensure that our initial surface $x^0=0$ is spacelike w.r.t. $g$. Hence, by continuity, surfaces of constant $x^0$ are spacelike at least for small $x^0$. Define $3$ matrices
\be
\label{def_ABC_2}
 A^{IJ} = {\cal P}^{IJ00} \quad B^{IJ} = 2\xi_i {\cal P}^{IJ0i} \quad C^{IJ} = \xi_i \xi_j {\cal P}^{IJij}
\ee
If surfaces of constant $x^0$ are non-characteristic then $A^{IJ}$ is invertible and we can define the $22 \times 22$ matrix
\be
\label{Mdef}
 M(\xi_i) = \left( \begin{array}{cc} 0 & I \\ -A^{-1} C(\xi_i) & -A^{-1} B(\xi_i) \end{array} \right)  
\ee
Let $G^{ij}$ be a smooth (inverse) Riemannian metric on these surfaces. Strong hyperbolicity is the statement that, for any (real) unit (w.r.t $G^{ij}$) covector $\xi_i$ on such a surface, the matrix $M(\xi_i)$ admits a {\it symmetrizer}: a positive definite hermitian matrix $K(\xi_i)$ such that
$K(\xi_i) M(\xi_i) = M(\xi_i)^\dagger K(\xi_i)$. $K(\xi_i)$ must depend {\it smoothly} on $\xi_i$ and also on the spacetime coordinates $x^\mu$ that we have suppressed above. Strong hyperbolicity implies that $M(\xi_i)$ is diagonalizable with real eigenvalues. Conversely, strong hyperbolicity follows if $M(\xi_i)$ is diagonalizable with real eigenvalues and eigenvectors depending smoothly on $\xi_i$. $\xi_0$ is an eigenvalue of $M(\xi_i)$ iff $\xi_\mu = (\xi_0,\xi_i)$ is a characteristic covector. The corresponding eigenvectors have the form $(T_I,\xi_0 T_I)^T$ where $T_I$ satisfies \eqref{characteristic}. 

Consider first the 2-derivative ($2\partial$) theory obtained by setting $\alpha=\beta=0$ in \eqref{4dST}. In this case, $\xi_\mu$ is characteristic if, and only if, it is null w.r.t. one of our three metrics \cite{newpaper}. This implies that spacelike (w.r.t. $g$) surfaces are non-characteristic. For given $\xi_i$ there are two characteristics null w.r.t. each metric (associated to the future and past null cones), and two corresponding real eigenvalues $\xi_0$. Hence $M(\xi_i)$ has $6$ real eigenvalues. The characteristics $\tilde{\xi}_\mu^\pm = (\tilde{\xi}_0^\pm,\xi_i)$ null w.r.t. $\tilde{g}^{\mu\nu}$ arise from a residual gauge symmetry of \eqref{def_har}: each has a 4d space of solutions of \eqref{characteristic} and hence a 4d eigenspace of $M(\xi_i)$. The two characteristics $\xi_\mu^\pm = (\xi_0^\pm,\xi_i)$ null w.r.t. $g^{\mu\nu}$ are associated with physical polarizations, each with a 3d eigenspace (corresponding to 2 graviton and 1 scalar field degree of freedom). The two characteristics $\hat{\xi}_\mu^\pm = (\hat{\xi}_0^\pm,\xi_i)$ null w.r.t. $\hat{g}^{\mu\nu}$ are also characteristics of \eqref{Heq}; these are associated with ``gauge-condition violating" polarizations. Each has a 4d eigenspace. The total dimensionality of the eigenspaces is $22$ so $M(\xi_i)$ is diagonalizable with real eigenvalues $\xi_0$. In each case the solutions $T_I$ of \eqref{characteristic} depend smoothly on $\xi_i$ \cite{newpaper}. This is sufficient to ensure strong hyperbolicity. Thus our modified harmonic gauge formulation of the $2\partial$ theory admits a well-posed initial value formulation. By setting the scalar field to zero, this new formulation also applies to vacuum GR.

We now include the 4-derivative ($4\partial$) terms. Decompose the principal symbol into a part ${\cal P}_2(\xi)$ arising from the $2\partial$ terms in \eqref{def_har} (including the gauge-fixing terms) and a part $\delta {\cal P}(\xi)$ arising from the $4\partial$ terms. Explicit expressions for the latter can be found in \cite{Papallo:2017ddx}. By ``weakly coupled" we mean that the components $\delta {\cal P}^{IJ \mu\nu}$ are small compared to ${\cal P}_2^{IJ\mu\nu}$. This will be the case if the components of the Riemann tensor, and the first and second derivatives of the scalar field, are small compared to any length scales (e.g. coupling constants) appearing in the higher-derivative terms. By continuity, if initial data is chosen so that the theory is weakly coupled then the resulting solution will be weakly coupled at least for a small time interval. 

Spacelike surfaces of constant $x^0$ are non-characteristic iff $\det A^{IJ} \ne 0$. This condition is satisfied in the $2\partial$ theory and so, by continuity, it is also satisfied in the $4\partial$ theory at sufficiently weak coupling. However, this condition may fail at strong coupling. 
The eigenvalues of $M(\xi_i)$ depend continuously on $M(\xi_i)$ and so, at weak coupling, we can divide the eigenvalues into $6$ groups according to which eigenvalue ($\tilde{\xi}_0^\pm,\xi_0^\pm$ or $\hat{\xi}_0^\pm$) of the $2\partial$ theory they reduce to at zero coupling. For each group we can define a ``total generalized eigenspace" as the direct sum of the spaces corresponding to the Jordan blocks of the eigenvalues in that group \cite{Papallo:2017qvl}. This defines $6$ (complex) vector spaces which we denote $\tilde{V}^\pm$, $V^\pm$ and $\hat{V}^\pm$. 

The ``pure gauge" characteristics $\tilde{\xi}_\mu^\pm$ of the $2\partial$ theory are also characteristics of the $4\partial$ theory. Hence $\tilde{\xi}_0^\pm$ are eigenvalues of $M(\xi_i)$. The eigenvectors are the same as for the $2\partial$ theory. Thus $\tilde{V}^\pm$ are 4d genuine eigenspaces. A continuity argument \cite{newpaper} establishes that, at weak coupling, the covectors $\hat{\xi}^\pm_\mu$ are also characteristic so $\hat{\xi}_0^\pm$ are eigenvalues of $M(\xi_i)$. Each is associated with $4$ eigenvectors that depend smoothly on $\xi_i$. So $\hat{V}^\pm$ are also 4d genuine eigenspaces. (In standard harmonic gauge this argument fails because the ``pure gauge" and ``gauge-condition violating" eigenvalues are degenerate with each other. This allows the matrix $M(\xi_i)$ to develop non-trivial Jordan blocks when one deforms from the $2\partial$ to the $4\partial$ theory \cite{Papallo:2017qvl,Papallo:2017ddx}. Our modified harmonic gauge formulation eliminates this degeneracy and thereby avoids this problem.)

The spaces $V^\pm$ are associated to the ``physical" eigenvalues. In this case, we expect the 3-fold degeneracy of the $2\partial$ theory to be split by the $4\partial$ terms, i.e., generically the two graviton polarizations and the scalar field will propagate with different speeds. In this case, it is not clear that the associated eigenvectors of $M(\xi_i)$ will depend smoothly on $\xi_i$ at values for which degeneracy of eigenvalues occurs. To evade this problem we construct a symmetrizer directly. Consider the matrices (our sign convention is $\mp g^{0\mu} \xi^\pm_\mu>0$)
\be\label{defHstar}
  H_\star^\pm = \pm \left( \begin{array}{ll} B_\star & A_\star \\ A_\star & 0 \end{array} \right)
\ee
where $A_\star$ and $B_\star$ are defined as in \eqref{def_ABC_2} but omitting the gauge-fixing terms from ${\cal P}$. These matrices are real symmetric \cite{Papallo:2017qvl}. Define a Hermitian form on $V^\pm$ by $(v^{(1)},v^{(2)})_\pm = v^{(1)\dagger} H_{\star}^\pm v^{(2)}$ where $v^{(1)},v^{(2)} \in V^\pm$. It can be shown that $(,)_\pm$ is positive definite in the $2\partial$ theory \cite{newpaper}. Hence, by continuity, it is positive definite at weak coupling in the $4\partial$ theory, and therefore defines an inner product on $V^\pm$.

It can be shown \cite{newpaper} that $H_\star^\pm$ is a symmetrizer for $M(\xi_i)$ within $V^\pm$. In particular, this implies that the eigenvalues associated with $V^\pm$ are real, and that $V^\pm$ admits a basis of eigenvectors. The latter may fail to be smooth in $\xi_i$ at points of degeneracy. But the symmetrizer $H_\star^\pm$ is smooth by definition. A symmetrizer for $M(\xi_i)$ can now be constructed as a block diagonal matrix where the blocks associated to $V^\pm$ are $H_\star^\pm$ and the blocks associated to the other spaces are constructed from the (smooth) eigenvectors on these spaces in the usual way.


\section{Discussion}

Several steps in our argument make use of the weakly coupled assumption. If the theory enters a strongly coupled regime then well-posedness can fail \cite{Papallo:2017qvl,Ripley:2019hxt,Ripley:2019irj,Ripley:2019aqj} but, from an EFT perspective, we do not expect  \eqref{4dST} to be valid at strong coupling anyway. 

Although we have focused on the theory \eqref{4dST}, our modified harmonic gauge condition can be applied to obtain strongly hyperbolic formulations of any weakly coupled Lovelock \cite{Lovelock1971} or Horndeski \cite{Horndeski1974} theory \cite{newpaper}. The former includes Einstein-Gauss-Bonnet theory, which gives the leading four-derivative EFT corrections to vacuum GR in higher dimensions.  

We saw above that, given a solution $(M,g)$ of \eqref{def_har} arising from initial data satisfying the constraint equations and gauge condition on $\Sigma \subset M$, this solution will satisfy \eqref{eofm} throughout $D(\Sigma)$. We define ${\cal D}(\Sigma)$ to be the region in which the solution is {\it uniquely determined} by the initial data. For the $2\partial$ theory we will have ${\cal D}(\Sigma) = D(\Sigma)$. But for a weakly coupled $4\partial$ theory, generically, some of the physical characteristics will be spacelike w.r.t. $g$. Since information can propagate along these characteristics, this will imply ${\cal D}(\Sigma) \subset D(\Sigma)$. Our analysis establishes {\it local} well-posedness, which ensures uniqueness in a neighbourhood of $\Sigma$.

Our formulation depends on the choice of the auxiliary metrics $\tilde{g}^{\mu\nu}$ and $\hat{g}^{\mu\nu}$. One way of choosing these is to set $\tilde{g}^{\mu\nu} = g^{\mu\nu}- a n^\mu n^\nu$ and $\hat{g}^{\mu\nu} =  g^{\mu\nu} - b n^\mu n^\nu$ where $n^\mu$ is a unit (w.r.t. $g$) vector field and $a(x),b(x)$ are functions. In a numerical simulation one might choose $n^\mu$ to be normal to surfaces of constant $x^0$ and $a,b$ to be constants. The ordering of the $3$ null cones assumed above requires $0<a<b$. However, this ordering can be changed as long as the null cones do not intersect and surfaces of constant $x^0$ are spacelike w.r.t. to all three metrics \cite{newpaper}. Such a change would affect the domain of dependence properties of the equation. 

Finally, our modified harmonic gauge may be useful even in conventional GR. We will discuss this in \cite{newpaper}.

{\it Acknowledgments}. We thank F. Abalos, P. Figueras and O. Reula for helpful conversations. ADK is supported by the George and Lilian Schiff Studentship. HSR is supported by STFC Grants PHY-1504541 and ST/P000681/1.


\end{document}